\documentclass[11pt]{article}
\usepackage{times}
\usepackage{mathfont}

\input epsf
\def\efig#1#2{\hbox{\epsfxsize=#1\epsfbox{#2}}}

\DeclareSymbolFont{AMSb}{U}{msb}{m}{n}
\DeclareSymbolFontAlphabet{\Bbb}{AMSb}

\def\:{\kern 0.1em}
\def\max{\mathop{\rm max}}
\def\log{\mathop{\rm log}}

\newtheorem{theorem}{Theorem}
\newtheorem{lemma}{Lemma}
\newtheorem{corollary}{Corollary}
\newtheorem{defn}{Definition}

\DeclareSymbolFont{lasy}{U}{lasy}{m}{n}
\let\Box\undefined
\DeclareMathSymbol\Box{0}{lasy}{"32}
\newcommand{\qed}{~$\Box$\medbreak}
\newenvironment{proof}{\noindent{\bf Proof: }}{\qed}

\newcommand{\paramvec}{{\rm\bf p}}
\def\cost#1{{\rm\bf c}_{#1}}

\makeatletter
\long\def\@makecaption#1#2{
   \vskip 10pt 
   \setbox\@tempboxa\hbox{{\small #1. #2}}
   \ifdim \wd\@tempboxa >\hsize   
       {\small #1. #2}\par        
     \else                        
       \hbox to\hsize{\hfil\box\@tempboxa\hfil}  
   \fi}

\def\@begintheorem#1#2{\it\trivlist
  \item[\hskip\labelsep{\bf #1\ #2.\ }]}
\def\@opargbegintheorem#1#2#3{\it\trivlist
  \item[\hskip\labelsep{\bf #1\ #2\ {\rm(#3)}.}]}
\makeatother

\let\setminus -

\begin{document}
\title{Setting Parameters by Example}
\author{David Eppstein
\thanks{Dept. Inf. \& Comp. Sci.,
Univ. of Calif., Irvine,
CA 92697-3425,
eppstein@ics.uci.edu,
http://www.ics.uci.edu/$\sim$eppstein/}}
\date{}
\maketitle

\begin{abstract}
We introduce a class of ``inverse parametric optimization''
problems, in which one is given both a parametric optimization problem
and a desired optimal solution; the task is to determine parameter
values that lead to the given solution.  We describe algorithms for
solving such problems for minimum spanning trees, shortest paths, and
other ``optimal subgraph'' problems, and discuss applications in
multicast routing, vehicle path planning, resource allocation, and board
game programming.
\end{abstract}

\section{Introduction}

Many cars now come equipped with route planning software that suggests
a path from the current location to a desired destination.
Similar services are also available on the internet (e.g., from
http://maps.yahoo.com/).  But although these routes may be found by
computing shortest paths in a graph representing the local road system, the
``distance'' may be a weighted sum of several values other than actual
mileage: expected travel time, scenic value, number of turns, tolls,
etc.~\cite{RogLan-AAAI-98}. Different drivers may have different
preferences among these
values, and may not be able to clearly articulate these preferences.
Can we automatically infer the appropriate weights to use in the sum by
observing the routes actually chosen by a driver?

More abstractly, we define an {\em inverse parametric optimization}
problem as follows: we are given as input both a {\em parametric
optimization} problem (that is, a combinatorial optimization problem
such as shortest paths, but with the element weights being linear
combinations of certain parameters rather than fixed numbers), and also
a desired optimal solution for the problem.\footnote{One could more
generally allow as input a set of problem-solution pairs, but for most
of the problems we consider any such set can be represented equally well
by a single larger problem.}  Our task is to determine parameter values such
that the given solution is optimal for those values.

Along with the path planning problem described above,
one can find many other applications in which one must tune
the parameters to an optimization problem:

\begin{itemize}
\item In many online services such as web page hosting,
data is sent in a star topology from a central server
to each user.  But
in multicast routing of video and other high-bandwidth
information, network resources are conserved by sending the
data along the edges of a tree,
in which some users receive copies of the data from other
users rather than from the central server.
Natural measures of the quality of each edge in
this routing tree include the edge's bandwidth, congestion,
delay, packet loss, and possibly monetary charges for use of that link.
Since one can find minimum spanning trees efficiently in the distributed
setting~\cite{GarKutPel-SJC-98}, it is natural to try to model this
routing problem using minimum spanning trees.
Given one or more networks with these parameters, and
examples of desired routing trees, how can we set the
weights of each quality measure
so that the desired trees are the minimum spanning trees of their networks?

\item Bipartite matching, or the {\em assignment problem}, is a common
formalism for grouping indivisible resources with resource consumers.
For instance, the first example given for matching by Ahuja et
al.~\cite{AhuMagOrl-93} is to assign recently hired workers to jobs,
using weights
based on such values as aptitude test scores and college grades.
One might set the weight of an edge
from worker $i$ to job $j$ to be $a_i\cdot p_j$,
where $a_i$ is the (known) set of aptitudes of the worker,
and $p_j$ is the (unknown) set of parameters describing
the combination of aptitudes best fitting the job.
Again, it is natural to ask for a way to automatically set the parameters
of each job, based on experience assigning previously hired workers to
those jobs.

\item Many board games, such as chess, checkers, or Othello, can
be played well by programs based on relatively simple
alpha-beta searching algorithms. 
However, these programs use relatively complex evaluation functions
in which the evaluation of a given position can be the sum of hundreds
or thousands of terms.  Some of these terms may represent the gross material
balance of a game (e.g., in chess, one usually normalizes the score so
that a pawn is worth 1 point, while a knight may be worth 2.5-3 points)
while others represent more subtle features of piece placement, king safety,
advanced pawns, etc.  The weight of each of these terms may be
individually adjusted in order to improve the quality of play.
Although there have been some preliminary experiments in using evolutionary
learning techniques to tune these weights~\cite{Sta-96},
they are currently usually set by hand.
The true test of a game program is in actual play, but programs
are also often tuned by using {\em test suites}, large collections
of positions for which the correct move is known.
If we are given a test suite,
can we automatically set evaluation weights in such a way that
a shallow alpha-beta search can find each correct move?
\end{itemize}

\subsection{New Results}

We show the following theoretical results:

\begin{itemize}
\item For the inverse parametric minimum spanning tree problem,
in the case that the number of parameters is a fixed constant,
we provide a randomized algorithm with linear expected running time, and
a deterministic algorithm with worst case running time $O(m\log^2 n)$.

\item For the minimum spanning tree, shortest path, matching, and other
``optimal subgraph''
problems for which the optimization problem can be solved in polynomial time,
we show that the inverse optimization problem can also be solved in
polynomial time by means of the ellipsoid method from linear programming,
even when the number of parameters is large.
\end{itemize}

In addition, although we do not provide theoretical results for this case,
we discuss the game tree search problem and describe how to fit it into
the same inverse parametric optimization framework.

In cases where the initial problem is infeasible
(there is no parameter setting leading to the desired optimum),
our techniques provide a witness for infeasibility:
a small number of alternative solutions,
one of which must be better than the given solution for any parameter
setting. One can then examine these solutions to determine whether the
initial solution is suboptimal or whether additional parameters should
be added to better model the users' utility functions.

\subsection{Relation to Previous Work}

Although there has been considerable work on
parametric versions of optimization problems such as
minimum spanning trees~\cite{AgaEppGui-FOCS-98}
and shortest paths~\cite{YouTarOrl-Nw-91},
we are not aware of any prior work in inverting such problems
to produce parameter values that match given solutions.
One could compute the set of solutions available over the range of
parameter values, and compare these solutions to the given one,
but the number of different solutions would typically grow exponentially
with the number of parameters.

The inverse parametric optimization problems considered here are most
closely related to {\em parametric search}, which describes a general
class of problems in which one sets the parameters of a parametric problem
in order to optimize some criterion.
However in most applications of parametric search, the criterion being
optimized is a numeric function of the solution (e.g. the ratio between
two linear weights) rather than the solution structure itself.
Megiddo \cite{Meg-JACM-83} describes a very general technique for solving
parametric search problems, in which one simulates the steps of an
optimization algorithm, at each conditional step using the algorithm itself as
an oracle to determine which conditional branch to take.
However this technique does not seem to apply to our problems, because
the given optimal structure (e.g. a single shortest path) does not give enough
information to deduce the conditional branches followed by a shortest
path algorithm.

The vehicle routing problem discussed in the introduction was introduced
by Rogers and Langley~\cite{RogLan-AAAI-98}.  However, they used a
weaker model of optimization (a hill-climbing procedure) and a stronger
model of user interaction requiring the user to specify preferences in a
sequence of choices between pairs of routes.

\section{Minimum Spanning Trees}

In this section, we consider the {\em inverse parametric minimum
spanning tree problem}, in which we are given a fixed tree $T$
in a network in which the weight of each edge $e$ is a linear function
$w(e)=\cost{e}\cdot\paramvec$
(where $\paramvec$ represents the unknown vector of parameter settings
and $\cost{e}$ represents the known value of edge~$e$ according to each
parameter).
Our task is to find a value of $\paramvec$ such that $T$
is the unique minimum spanning tree for the weights~$w(e)$.

If we fix a given spanning tree $T$ in a network, a pair of edges
$(e,f)$ is defined to be a {\em swap} if $T\cup\{f\}\setminus\{e\}$
is also a spanning tree;
that is, if $e$ is an edge in $T$, $f$ is not an edge in $T$,
and $e$ belongs to the cycle induced in $T$ by $f$.
$T$ is the unique minimum spanning tree if and only if for every swap $(e,f)$,
the weight of $f$ is greater than the weight of $e$.

Thus we can solve the inverse parametric minimum spanning tree problem
as a linear program, in which we have one variable per parameter,
and one constraint $(\cost{f}-\cost{e})\cdot\paramvec>0$ per swap.
If the number of variables is a fixed constant,
a linear program may be solved in time linear in the number of
constraints~\cite{Meg-JACM-84}; however here the number of constraints may
be $\Theta(m\:n)$.

We show how to improve this by a randomized algorithm which
takes linear time and a deterministic algorithm which
takes time $O(m\log^2 n)$.  Both algorithms are based on (different)
random sampling schemes for low dimensional linear programming,
due to Clarkson~\cite{Cla-JACM-95}.

\subsection{Randomized Spanning Tree Algorithm}

Clarkson \cite{Cla-JACM-95} showed that, if one randomly samples $k$ constraints
from a $d$-dimensional linear program with $n$ constraints, and computes
the optimum
for the subprogram consisting only of the sampled constraints,
then the expected number of the remaining constraints violated
by this optimum is at most $d(n-k)/(k+1)$.  Further, if any constraint is violated,
at least one of the $d$ constraints
involved in any {\em base} (minimal subset of constraints having the
same solution as the overall problem) belongs to the set of
violated constraints.  If no constraint is violated, the problem is solved.

This suggests the following randomized algorithm for the inverse
parametric minimum spanning tree problem,
where $d=O(1)$ is a fixed constant.
We define a {\em potential swap} for the given tree $T$ to be a
pair $(e,f)$ where $e$ belongs to $T$ and $f$ does not,
regardless of whether $(e,f)$ is actually a swap.
For technical reasons, we need to define a unique optimal parameter
setting $\paramvec$ for any subset of constraints, which we achieve by
introducing an arbitrary linear objective function.

\vfill\eject
\begin{enumerate}
\item Let set $S$ be initialized to empty.
\item Repeat $d$ times:
\begin{enumerate}
\item Let set $R$ be a random sample of $d\sqrt{m\:n}$ potential swaps.
\item Find the optimal parameter setting $\paramvec$ for constraints from $R\cup S$.
\item Add the constraints violated by $\paramvec$ to $S$.
\end{enumerate}
\item Find the optimal parameter setting $\paramvec$ for constraints from $S$.
\end{enumerate}

Each iteration increases the size of the intersection of $S$
with the optimal base, so the loop terminates with a correct solution.
The expected number of edges added to $S$
in each iteration is $O(\sqrt{m\:n})$,
so the expected size of $S$ is $O(d\sqrt{m\:n})=O(m)$.
If $d=O(1)$, the step in which we find $\paramvec$ can be performed
in time $O(d\sqrt{m\:n})=O(m)$ by fixed dimensional linear programming
techniques.
It remains to determine how we tell whether a potential swap $(e,f)$
is really a swap (so we can determine whether to use it as a constraint
or ignore it in step (b)), and how to find the set of violated
constraints (step (c)).

To test a potential swap, we simply build a least common ancestor data
structure \cite{SchVis-SJC-88} on the given tree $T$ (with an arbitrary
choice of root).  The pair $(e,f)$ is a swap if both
endpoints of $e$ are on the path from one of the endpoints of $f$
to the common ancestor of the two endpoints.

\begin{figure}[t]
$$\efig{4.5in}{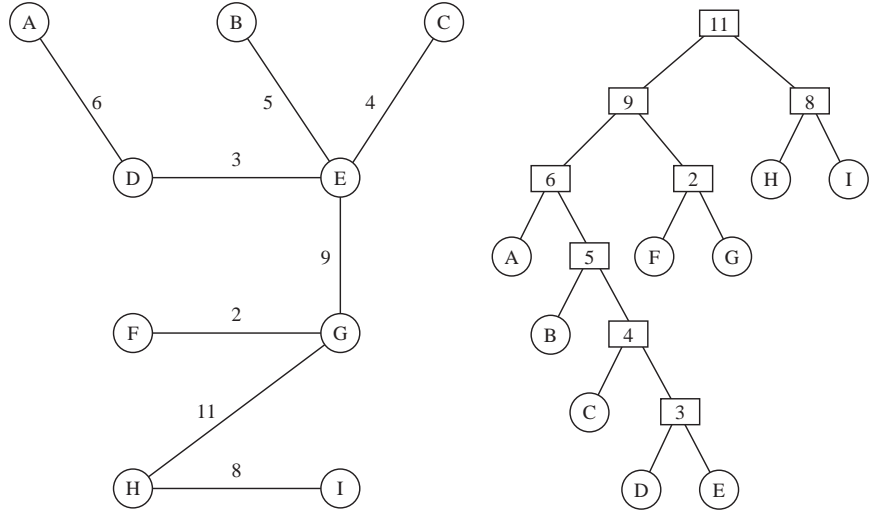}$$
\caption{(a) A weighted tree $T$; (b) Auxiliary tree used to find
heaviest edges on paths in $T$.}
\label{fig:heavy}
\end{figure}
\def\respartfig{\ref{fig:part}(a)}
\def\mlpartfig{\ref{fig:part}(b)}

To find the violated constraints for $\paramvec$, we also use least
common ancestors, on an auxiliary tree in which internal nodes represent
edges and leaves represent vertices of $T$ (Figure~\ref{fig:heavy}).  We
build this auxiliary tree by choosing the root to be the maximum weight
edge $e$ (according to
$\paramvec$) in $T$, with the two children of the root being auxiliary
trees constructed recursively on the two components of $T\setminus\{e\}$.
This construction takes time $O(n\log n)$.
The least common ancestor of two leaves in this auxiliary tree
represents the maximum weight edge on the path between the corresponding
vertices of $T$.  Therefore, if $f$ is a given non-tree edge, we
can find a swap $(e,f)$ giving a violated constraint (if one exists)
by using this auxiliary tree to find the maximum weight edge on the path
between $f$'s endpoints.  If this gives us a violated swap,
we continue recursively on the subpaths between $f$ and the endpoints of $e$,
until all swaps involving $f$ have been listed.
Each swap is found in $O(1)$ time, and the expected number of swaps
corresponding to violated constraints is $O(\sqrt{m\:n})$,
so the total expected time for this procedure
(including the time to construct the auxiliary tree)
is $O(m+n\log n)$.

\begin{lemma}\label{logrand}
We can solve the inverse parametric minimum spanning tree problem,
for any constant number of parameters, in randomized
expected time $O(m+n\log n)$.
\end{lemma}

In order to remove the unnecessary logarithmic factor from this bound,
we resort to another round of sampling.
However this time we sample tree edges rather than swaps.

\begin{lemma}\label{edge-sample}
Let $S$ be a randomly chosen sample of $k$ edges from tree $T$,
let graph $G'$ and tree $T'$ be formed from $G$ and $T$ respectively by
contracting the edges in
$T\setminus S$, and let $\paramvec$ be the optimal parameter setting
for the inverse parametric minimum spanning tree problem defined by $G'$
and $T'$.
Then the expected number of the remaining edges of $T$
that take part in a constraint violated
by this optimum is at most $d(n-k-1)/(k+1)$.
\end{lemma}

\begin{proof}
Consider selecting $S$ in the following way: choose a random permutation
on the edges of $T$, and let $S$ be the first $k$ edges in the permutation.
Let $e$ be the $(k+1)$st edge in the permutation.
Then since $e$ is equally likely to be any remaining edge,
the expected number of edges that take part in a violated constraint
is just $n-d-1$ times the probability that $e$ takes part in a violated
constraint.  But this can only happen if $e$ is one of the at most $d$
edges involved in the optimal base for $S\cup\{e\}$.  Since this subset
is just the first $k+1$ edges in the permutation, and
any permutation of this subset is equally likely, this probability
is at most $d/(k+1)$.
\end{proof}

Thus we can apply the following algorithm:

\begin{enumerate}
\item Let set $S$ be initialized to empty.
\item Repeat $d$ times:
\begin{enumerate}
\item Let set $R$ be a random sample of $d\sqrt{n}$ edges of $T$.
\item Contract the edges in $T\setminus (R\cup S)$ to produce $T'$ and $G'$.
\item Find the optimal parameter setting $\paramvec$ for $T'$ and $G'$
using the algorithm of Lemma~\ref{logrand}.
\item Add to $S$ the tree edges that take part in
a constraint violated by $\paramvec$.
\end{enumerate}
\item Contract the edges in $T\setminus S$ to produce $T'$ and $G'$.
\item Find the optimal parameter setting $\paramvec$ for $T'$ and $G'$
using the algorithm of Lemma~\ref{logrand}.
\end{enumerate}

The arguments for termination and correctness are the same as before.
It remains to explain how we find the set of violated tree edges.
This can be done in time $O(m\alpha(m,n))$ by an algorithm of
Tarjan~\cite{Tar-JACM-79}, but using this algorithm directly
would lead to a nonlinear overall time bound.
More recent minimum spanning tree verification algorithms
\cite{DixRauTar-SJC-92,Kin-Algo-97} can be used to find the violated non-tree
edges, but not the tree edges.  However, in our case we can perform this
verification task efficiently due to the expected small number of
differences between $T$ and the minimum spanning tree for $\paramvec$.

\begin{lemma}\label{verify}
In the algorithm above, the tree edges that take part in a violated
constraint can be found in expected linear time.
\end{lemma}

\begin{proof}
We use the linear time randomized minimum spanning tree algorithm of
Karger et al.~\cite{KarKleTar-JACM-95}, and let $X$ denote the set of edges
that are in the MST and not in $T$.
Note that $X$ has exactly as many edges as are in $T$ and not in
the MST; since each edge in
the latter set takes part in a violated swap constraint,
the expectation of $|X|$ is $O(\sqrt n)$
by Lemma~\ref{edge-sample}.
Then it is easy to see that, if tree edge $e$ takes part in any violated
constraints, at least one must be the constraint corresponding to swap $(e,f)$,
where $f$ is the minimum weight edge in $X$ forming a swap with $e$.

To find this minimum weight swap for each tree edge, we contract $T$ as
follows.  While $T$ has a degree-one vertex that is not adjacent to any
edge in $X$, we remove it and its incident edge; that edge can not take
part in any swaps with $X$.  While $T$ has a degree-two vertex that is
not adjacent to any edge in $X$, we remove it and merge its two incident
edges into a single edge; these two edges share the same minimum swap edge.

After this contraction process, the contracted tree $T'$ has $O(|X|)$
vertices with degree less than three, and therefore $O(|X|)$ total vertices.
We apply Tarjan's nonlinear minimum spanning tree verification algorithm
to this contracted tree to find the best swap in $X$ for each contracted
tree edge. We then undo the contraction process and propagate the best swap
information to the original tree edges.  Finally, once we have computed
the best swap $(e,f)$ for each tree edge $e$, we simply compute
$w(e)$ and $w(f)$ and compare the two weights to determine whether this
swap leads to a violated constraint.
\end{proof}

\begin{theorem}
We can solve the inverse parametric minimum spanning tree problem,
for any constant number of parameters, in randomized
linear expected time.
\end{theorem}

\begin{proof}
The problem is solved by the algorithm above.
In each iteration the expected size of the set added to $S$ is $O(\sqrt n)$,
so the total size of $R\cup S$ is $O(d\sqrt n)=O(\sqrt n)$.
In each iteration we add one more member of the optimal base to $S$,
so the algorithm terminates with the correct solution.
The steps in which we find the optimal parameter setting for $T'$ and $G'$
can be performed by applying Lemma~\ref{logrand};
since $T'$ has $O(\sqrt n)$ edges, 
the time for these steps is $O(m+\sqrt n\log n)=O(m)$.
The step in which we find the edges that take part in a violated constraint
can be performed in linear expected time by Lemma~\ref{verify}.
\end{proof}

\subsection{Deterministic Spanning Tree Algorithm}

To solve the inverse parametric minimum spanning tree problem
deterministically, we derandomize a different
sampling technique also based on a method of Clarkson~\cite{Cla-JACM-95}.
However, as in our randomized algorithm, we modify this technique
somewhat by sampling edges instead of constraints.

We begin by applying the {\em multi-level restricted partition}
technique of Frederickson~\cite{Fre-SJC-85,Fre-SJC-97} to the given tree
$T$.

\begin{figure}[t]
$$\efig{2.5in}{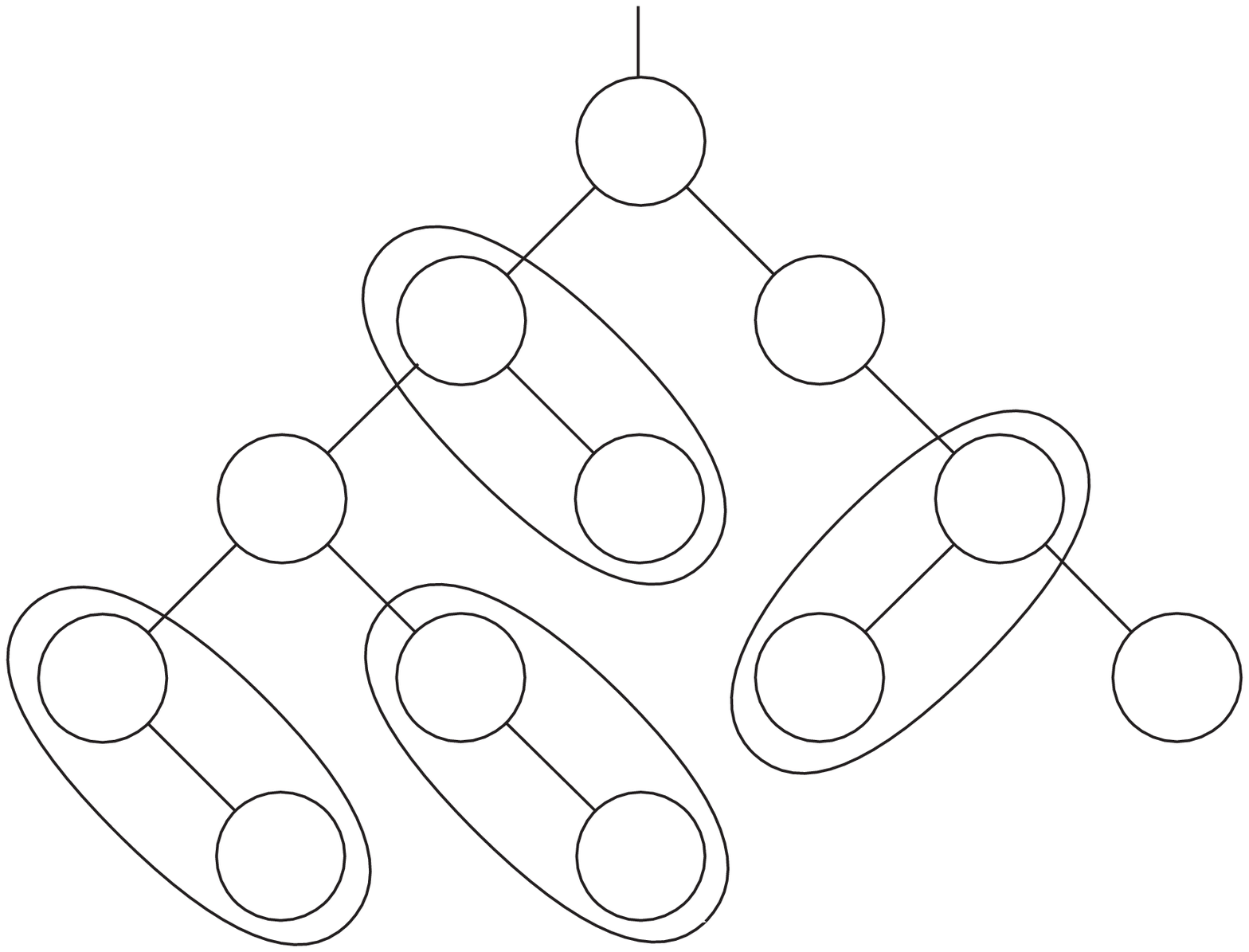}\qquad\qquad\quad
\efig{2.5in}{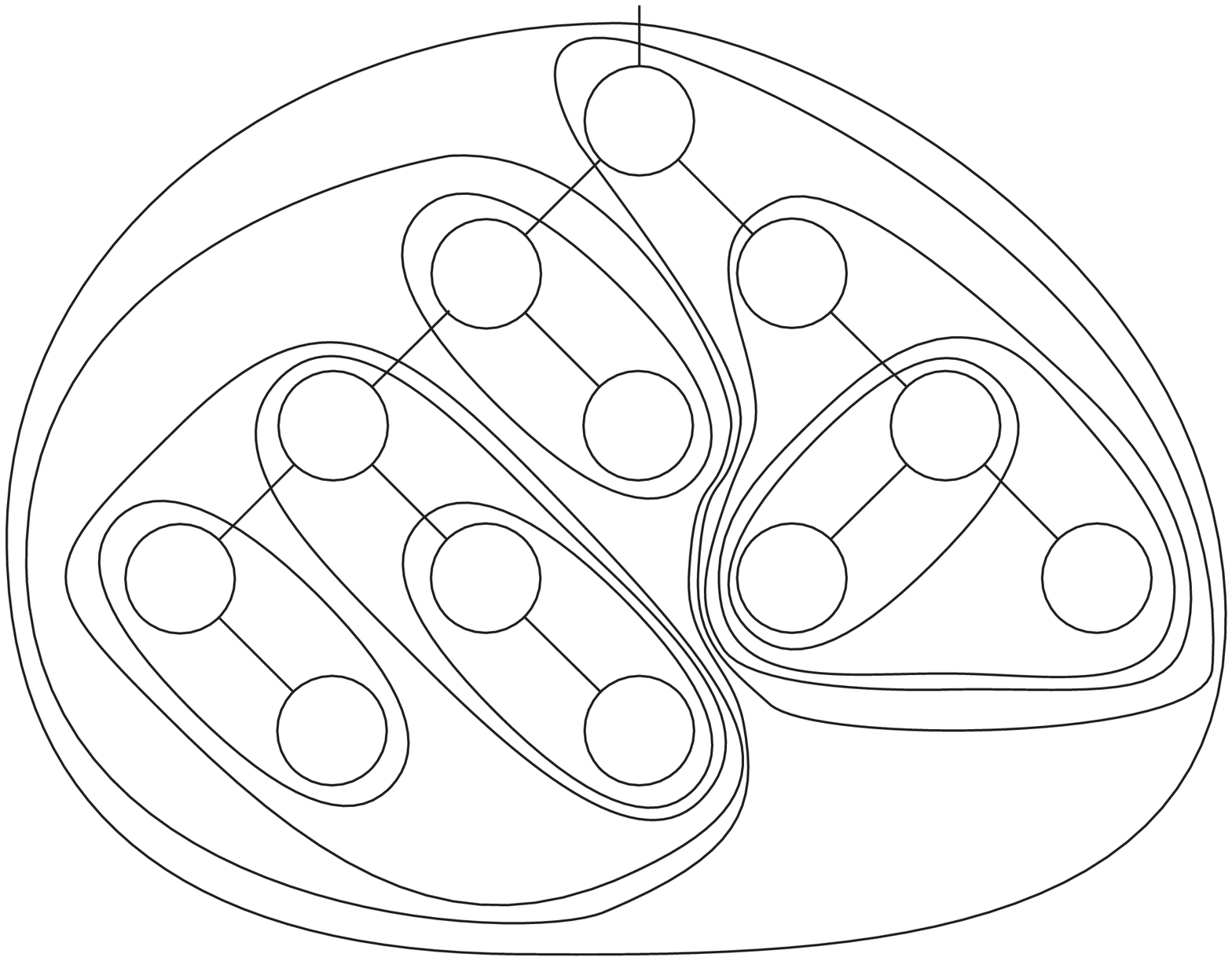}$$
\caption{(a) Restricted partition of order~2; (b) multi-level partition.}
\label{fig:part}
\end{figure}
\def\respartfig{\ref{fig:part}(a)}
\def\mlpartfig{\ref{fig:part}(b)}

By introducing dummy edges, we can assume without
loss of generality that $T$ is binary and that the root $t$ of $T$ has indegree
one.  These dummy edges will only be used to form the partition and will
not take part in the eventual optimization procedure.

\begin{defn}
A restricted partition of order $z$ with respect to a rooted binary tree $T$
is a partition of the vertices of $V$ such that:
\begin{enumerate}
\item Each set in the partition contains at most $z$ vertices.

\item Each set in the partition induces a connected subtree of $T$.

\item For each set $S$ in the partition,
if $S$ contains more than one vertex,
then there are at most two tree edges having one endpoint in $S$.

\item No two sets can be combined and still satisfy the other conditions.
\end{enumerate}
\end{defn}

Such a partition (for $z=2$) is depicted in Figure~\respartfig.
In general such a partition can easily be found in
linear time by merging sets until we get stuck.  Alternatively, by working
bottom up we can find an optimal partition in linear time. We will defer
until later choosing a value for $z$; for now we leave it as a free
parameter. 

\begin{lemma}[Frederickson \cite{Fre-SJC-97}]\label{few-sets}
Any order-$z$ partition of a binary tree $T$
has $O(n/z)$ sets in the partition.
For $z=2$ we can find a partition with at most $5n/6$ sets.
\end{lemma}

Contracting each set in a restricted partition gives again a binary
tree.  We form a {\em multi-level partition}~\cite{Fre-SJC-97} by
recursively partitioning this contracted binary tree (Figure~\mlpartfig).

We now use these partitions to construct a set $\Pi$ of paths in $T$.
We include in $\Pi$ the path in $T$ between any two vertices that are in
the same set at some level of the partition.
Note that, although the vertices at higher levels of the partition
correspond to contracted subtrees of $T$, the path in $T$ between two such
subtrees can still be unambiguously defined.

\begin{lemma}
The set of paths defined above has the following
properties:
\begin{itemize}
\item There are $O(n\:z)$ paths.
\item Each edge in $T$ belongs to $O(z^2\log_z n)$ paths.
\item Any path in $T$ can be decomposed into the disjoint
union of $O(\log_z n)$ paths.
\end{itemize}
\end{lemma}

\begin{proof}
The first property follows immediately from Lemma~\ref{few-sets},
since each set of the partition contributes $O(z^2)$ paths,
there are $O(n/z)$ sets at the bottom level of the partition, and
the number of sets decreases at least geometrically at each level. 
Similarly, the second property follows, since an edge can belong to
$O(z^2)$ paths per level and there are $O(\log_z n)$ levels.

Finally, to prove the third property, let $p$ be an arbitrary path in
$T$.  We describe a procedure for decomposing $p$ into few paths
$\pi_i\in\Pi$.  More generally, suppose we have a path $p$ contained in
a set $S$ at some level of a multi-level decomposition (note that the
whole tree is the set at the highest level of the partition).
Then $S$ can be decomposed into at most $z$ sets at the next level of
the partition; $p$ has endpoints in at most two of these sets, and
may pass completely through some other sets.  Therefore, $p$ can be
decomposed into the union of two smaller paths in the sets containing
its endpoints, together with a single path $\pi_i$ connecting those two
sets.  By repeating this decomposition recursively at
each level of the tree, we obtain a decomposition into at most two paths
per level, or $O(\log_z n)$ paths overall.
\end{proof}

We now describe how to use this path decomposition in our inverse
optimization problem.
For each path $\pi_i\in\Pi$, let $A_i$ denote the
set of edges in $T$ belonging to $\pi_i$, and let $B_i$ denote the set of
edges in $G\setminus T$ such that $\pi_i$ is part of the decomposition
of the tree path between each edge's endpoints.
The total size of all the sets $A_i$ and $B_i$ is $O((m+n\:z^2)\log_z n)$,
and all sets can be constructed in time linear in their total size.

A pair $(e,f)$ is a swap if and only if
there is some $e$ for which $e\in A_i$ and $f\in B_i$.
With this decomposition,
the inverse parametric minimum spanning tree problem becomes equivalent
to asking for a parameter $\paramvec$ such that, for each $i$,
the weight of every member of $A_i$ is less than the weight of every
member of $B_i$.

For a single value of $i$, one could solve such a problem by
a $(d+1)$-dimensional linear program in which we augment the parameters
by an additional variable that is constrained to be greater than each
$e\in A_i$ and less than each $f\in B_i$, however adding a separate
variable for each $i$ would make the dimension nonconstant.

\begin{figure}
$$\efig{5in}{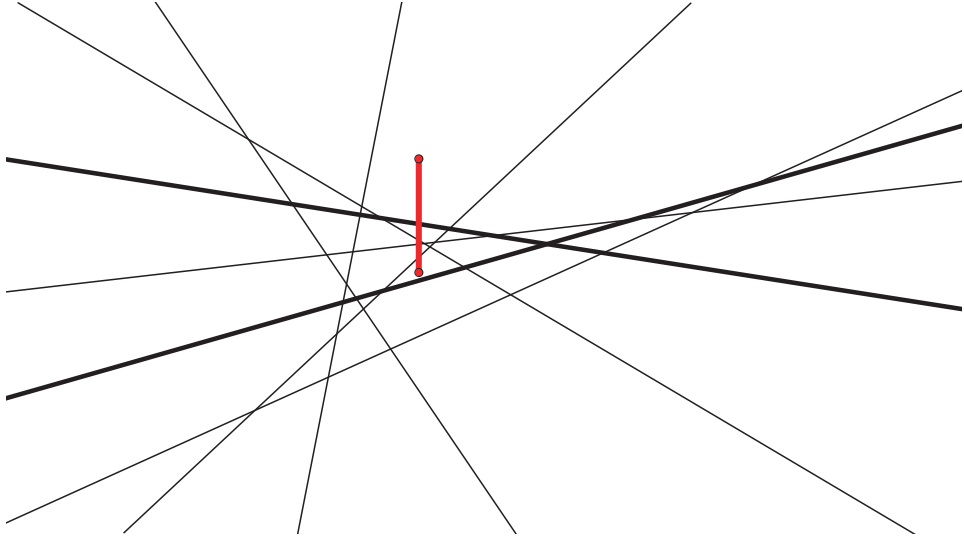}$$
\caption{Example $\epsilon$-net, for $\epsilon=1/2$: every vertical line
segment that crosses $\ge n/2$ lines in the overall arrangement also
crosses at least one of the two heavy lines.}
\label{epsilon}
\end{figure}

Instead, we use a standard derandomization technique from
computational geometry, {\em $\epsilon$-nets}.
If we graph the weight of each edge in a $(d+1)$-dimensional space,
where the parameter values are independent variables and the weight
is the dependent variable, the result is a hyperplane.
For any set $S$ of these hyperplanes, and any $\epsilon>0$,
define an $\epsilon$-net for vertical line segments
to be a subset $S'$ such that, if any vertical line segment
intersects at least $\epsilon |S|$ hyperplanes in $S$,
the same segment must intersect at least one hyperplane in $S'$
(Figure~\ref{epsilon}).
More generally, if the members of $S$ are given costs,
an $\epsilon$-net must contain at least one member of any
subset that is formed by intersecting the hyperplanes with
a vertical segment and that has total cost at least $\epsilon$
times the total cost of $S$.
If $1/\epsilon=O(1)$, an $\epsilon$-net of size $O(1)$
can be found in time linear in~$|S|$~\cite{Mat-JACM-95}.

Our algorithm can then be described as follows.
We will use $\epsilon=1/3d$.

\begin{enumerate}
\item Use a recursive partition to find the sets $A_i$ and $B_i$.
\item Assign unit cost to each edge in the graph.
\item Repeat until terminated:
\begin{enumerate}
\item Construct $\epsilon$-nets $A'_i$ and $B'_i$ for each $A_i$ and $B_i$.
\item Let $S$ be the set of swaps involving only $\epsilon$-net members.
Find the optimal parameter setting $\paramvec$ for constraints from $S$.
\item Find the maximum weight $a_i$ of an edge in each $A_i$ and the
minimum weight $b_i$ of an edge in each $B_i$,
where weights are measured according to~$\paramvec$.  If $a_i<b_i$
for each $i$, terminate the algorithm.
\item Find the maximum weight $a'_i$ of an edge in each $A'_i$ and the
minimum weight $b'_i$ of an edge in each $B'_i$.
Double the cost of each edge in $A_i$ with $w(e)>a'_i$,
and each edge in $B_i$ with $w(e)<b'_i$.
\end{enumerate}
\end{enumerate}

The set of edges in $A_i$ for which the costs are doubled
is defined by the intersection of $A_i$ with a vertical
line segment: the segment with parameter coordinates $\paramvec$
and with weight coordinate beween $a'_i$ and~$\infty$.
It does not contain any member of $A'_i$, so it must have
total cost at most $\epsilon$ times the cost of~$A_i$.
Therefore each iteration increases the total cost of all the sets $A_i$ (and
similarly $B_i$) by a factor of at most $1+\epsilon=1+1/3d$.

If there is any constraint violated by the solution $\paramvec$,
then at least one violated constraint must be a member of
the $d$-swap base defining the optimal overall solution.
Note however that, in any iteration of the loop, $a'_i<b'_i$
because of how we computed $\paramvec$,
so any violated constraint coming from a swap $(e,f)$ must
have $w(e)>a'_i$ or $w(f)<b'_i$.
Therefore
at least one of the $2d$ edges involved in the optimal base
must have its cost doubled,
and the cost of the optimal base increases by a factor of at least
$1+1/2d$.

Since the base's cost increases at a rate faster than the total cost,
it can only continue to do so for $O(d\log n)$ iterations
before it overtakes the total cost, an impossibility.
So at some point within those $O(d\log n)$ iterations
the algorithm must terminate the loop.

\begin{theorem}
We can solve the inverse parametric minimum spanning tree problem,
for any constant number of parameters, in worst case
time $O(m\log n\log_{m/n}n)$.
\end{theorem}

\begin{proof}
We use the algorithm described above, setting $z=\max(2,\sqrt{m/n})$.
Therefore, the total size of the sets $A_i$ and $B_i$
(and the total time to find these sets and to perform each iteration)
is $O(m\log_{m/n}n)$.
Since $d$ is constant, there are $O(\log n)$ iterations,
and the total time is $O(m\log n\log_{m/n}n)$.
\end{proof}

\section{Other Optimal Subgraph Problems}

We now describe a method for solving inverse parametric optimization on
a more general class of {\em optimal subgraph problems}, in which we are
given a graph with parametric edge weights and must find the minimum
weight {\em suitable subgraph}, where suitability is defined according
to the particular problem.  The minimum spanning tree problem considered
earlier has this form, with the suitable subgraphs simply being trees.
The shortest path and minimum weight matching problems also have this
form.  In order to solve these problems, we resort to the ellipsoid
method from linear programming.  This has the disadvantage of being not
strongly polynomial nor very practical, but its advantages are in its
extreme generality -- not only can we handle any optimal subgraph
problem for which the optimization version is polynomial, but (unlike
our MST algorithms) we are not limited to a fixed number of parameters.

A good introduction to the ellipsoid method and its applications in
combinatorial optimization can be found in the book by
Gr\"otschel, Lov\'asz, and Schrijver~\cite{GroLovSch-88}.

\begin{lemma}[Gr\"otschel, Lov\'asz, and Schrijver
\cite{GroLovSch-88}, p.~158]
For any
polyhedron $P$ defined by a strong separation oracle, and any rational
linear objective function $f$,
one can find the point in $P$ maximizing $f$
in time polynomial in the dimension of $P$ and in the maximum encoding
length of the linear inequalities defining $P$.
\end{lemma}

The {\em strong separation oracle} required by this result
is a routine that takes as input a $d$-dimensional point
and either determines that the point is in $P$ or returns
a closed halfspace containing $P$ and not containing the test point.
One slight technical difficulty with this approach is that it requires
the polyhedron to be closed (else one could not separate it from
a point on one of its boundary facets) while our problems
are defined by strict inequalities forming open halfspaces.
To solve this problem, we introduce an additional parameter $\delta$
measuring the separation of the desired optimal subgraph from other subgraphs,
and attempt to maximize $\delta$.

\begin{theorem}
Let $(G,X)$ be an inverse parametric optimization problem in which
$G$ is a graph with parametric edge weights, $X$ is the given
solution for an optimal subgraph problem, and there exists a polynomial
time algorithm that either determines that $X$ is the unique optimal subgraph
or finds a different optimal subgraph $Y$.
Then we can solve the inverse parametric optimization problem
for $(G,X)$ in time polynomial in the number of parameters,
in the size of the graph, and in the maximum encoding length
of the linear functions defining the edge weights of $G$.
\end{theorem}

\begin{proof}
We define a polyhedron $P$ by linear inequalities
$w(X) \le w(Y) - \delta$ where $w$ denotes the weight of a subgraph
for the given point $\paramvec$, $Y$ can be any suitable subgraph,
and $\delta$ is an additional parameter.  To avoid problems with
unboundedness, we can also introduce additional normalizing inequalities
${\bf -1}\le\paramvec\le{\bf 1}$.  Clearly, there exists
a point $(\paramvec,\delta)$ with $\delta>0$
in $P$ if and only if $\paramvec$ gives a feasible solution to the inverse
parametric optimization problem.

Although there can be exponentially many inequalities,
we can easily define an oracle that either terminates the entire
algorithm successfully or acts as a strong separation oracle:
to test a point $(\paramvec,\delta)$,
simply compute the optimal subgraph $Y$ for the weights defined by $\paramvec$.
If $X=Y$, we have solved the problem.
If $w(X)\le w(Y)-\delta$, the point is feasible.
Otherwise, return the halfspace $w(X)\le w(Y)-\delta$.

Therefore, we can apply the ellipsoid method to find the point maximizing
$\delta$ on $P$.  If the method returns a point with $\delta>0$
or terminates early with $X=Y$, we must have solved the problem,
otherwise the problem must be infeasible.
\end{proof}

\begin{corollary}
We can solve the inverse parametric minimum spanning tree, shortest
path, or matching problems in time polynomial in the size of the given
graph and in the encoding length of its parametric weight functions.
\end{corollary}

As a variant of this result, by using an algorithm
for finding the {\em second} best subgraph,
we can complete the ellipsoid method without early termination
and find a parameter value for which $X$ is optimally separated
from other subgraphs.
Efficient second-best algorithms are known for
minimum spanning trees \cite{DixRauTar-SJC-92,KatIbaMin-SJC-81,Kin-Algo-97},
shortest paths \cite{HofPav-JACM-59},
and matching \cite{Mih-PCIT-79};
in general the second-best subgraph
is the best subgraph within all graphs formed by deleting one edge of $X$
from~$G$.

\section{Game Tree Search}

As described in the introduction, we would like to be able to tune the
weights of a game program's evaluation function so that a shallow search
(to some fixed depth $D$) makes the correct move for each position
in a given test suite.  However, because of the possibility of making
the right move for the wrong reasons, this problem seems to be highly
nonlinear.  So, in order to apply our inverse parametric optimization technique
to this problem, we need some further assumptions.

Define an {\em unavoidable set} of positions for a given player and
depth $D$ to be a set of positions, each of which occurs $D$ half-moves
from the present situation, such that, no matter what the opponent does,
the given player can force the game to reach some position in the set.
More generally, we can define an unavoidable set for any subset of
positions to be a set such that, if the game ends within that subset,
the player on move can force it to be in the unavoidable set.  For any given
position, one can prove that one particular move is best by exhibiting
an unavoidable set $A_i$ for the positions reachable from that move
(from the perspective of the player to move) and an unavoidable set
$B_i$ (from the perspective of the other player) for the positions
reachable from the other moves, such that the minimum evaluation of any
position in $A_i$ is greater than the maximum evaluation of any position
in $B_i$.  Minimax or alpha-beta search can be interpreted as finding
both of these sets.

For a given position in a test suite, we will assume that
the position can be solved correctly by searching sufficiently deeply:
that is, there exists a depth $D'>D$ such that, if we search (with some
untuned or previously-tuned evaluation function) to depth $D'$, we will
find the correct move,
and not only that but we will find a correct depth-$D$ strategy:
unavoidable sets $A_i$ and $B_i$ at depth $D$
such that any good evaluation function should evaluate
all positions in $A_i$ greater than all positions in $B_i$.
We will therefore say that an evaluation function {\em evaluates the
position correctly}
if it evaluates all positions in $A_i$ greater than all positions in $B_i$.
If it does (and it implements a correct minimax search routine),
it must make the correct move in the given position.

Thus, the problem of finding an evaluation function that evaluates each
test suite position correctly can be cast into the same form used
in the deterministic minimum spanning tree algorithm:
a family of sets $A_i$ and $B_i$, and a requirement that the parameter choice
correctly sort the members of $A_i$ from the members of $B_i$.
However, there are two problems with using the $\epsilon$-net based
sampling approach of that algorithm.
First, the game evaluation problem seems likely to have many more parameters
than the minimum spanning tree problem, casting into doubt the
requirement that the number of parameters be a fixed constant.
And second, doing a deep search to compute and store the unavoidable sets
for each test suite position could be very costly.

Instead, we take the same approach used for the other optimal subgraph
problems, of using the ellipsoid method for linear programming with a
separation oracle.  In this case, the separation oracle consists of
running a depth-$D$ search on each test position, until one is found
at which the wrong move is made.
Once that happens, we can compute $A_i$ and $B_i$ for that one position,
using a deep search, and compare the values of the evaluation function
on those sets. (In fact the unavoidable sets by which the shallow search
``proves'' that it has the correct move for its evaluation must intersect
$A_i$ and $B_i$ in at least one member, so we can do this comparison
by a single shallow search.)
If this separation oracle finds an $a\in A_i$ and $b\in B_i$
that have evaluations in the wrong order,
it returns a constraint that the evaluation of $a$
should be greater than the evaluation of $b$.
Otherwise, if it fails to find a separating constraint, we may still not
evaluate each position correctly, but we must make the correct move in
each position.

\begin{theorem}
If there exists a setting of weights for an evaluation function
that evaluates each position of a given test suite correctly,
then we can find a setting that makes each move correctly.
The algorithm for finding this setting performs
a polynomial number of iterations,
where each iteration makes at most one shallow search
on each position of the suite, together with a single deep search
on a single suite position.
\end{theorem}

\section{Conclusions}

We have discussed several problems of inverse parametric optimization,
provided general solutions to a wide class of optimal subgraph problems
based on the ellipsoid method, and faster combinatorial algorithms
for the inverse parametric minimum spanning tree problem.

One difficulty with our approach comes from infeasible inputs:
what if there is no linear combination of parameters that leads to
the desired solution?  Rogers and Langley~\cite{RogLan-AAAI-98}
observe a similar phenomenon in their vehicle routing experiments,
and suggest searching for additional parameters to use.
This search may be aided by the fact that
infeasible linear programs can be witnessed by a small number of mutually
inconsistent constraints: in the path planning problem,
we can find $d+1$ paths, one of which must be better than the given path
for any combination of known parameters.  Studying these paths may reveal the
nature of the missing parameters.
Alternatively,
a search for a linear programming solution
with few violated constraints~\cite{Mat-DCG-95}
may provide a parameter setting for which the user's chosen solution is
near-optimal.

A natural direction for future research is in dealing with nonlinearity.
Problems in which the solution weight
includes low-degree combinations of element weights
(as are used in game programming
to represent interactions between positional features)
may be dealt with by including additional parameters for each
such combination.  But what about problems in which the element
weights are nonlinear combinations of the parameters?
For instance, if the parameters are coordinates of points,
any problem involving comparisons of distances will
involve quadratic functions of those coordinates.
The question of finding coordinates such that a given tree
is the Euclidean minimum spanning tree of the points is known to be
NP-hard~\cite{EadWhi-Algo-96},
but if the points' coordinates depend only on a constant number of
parameters one can solve the problem in polynomial time.  Can the
exponent of this polynomial be made independent of the number of parameters?

It may be possible to extend our spanning tree methods
to other matroids.  E.g., transversal matroids
provide a formulation of bipartite matching
in which the weights are on the vertices of one side of the bipartition,
rather than the edges.  Can we solve inverse parametric transversal
matroid optimization efficiently?  Are there natural applications of
this or other matroidal problems?

Another open question concerns the existence of combinatorial
algorithms for the inverse parametric shortest path problem.  It is
unlikely that a strongly
polynomial algorithm exists without restricting the dimension:
one can
encode any linear programming feasibility problem as an inverse
parametric shortest path (or other optimal subgraph) problem, by using a
parallel pair of edges for each constraint.
But is there a strongly polynomial algorithm for inverse parametric
shortest paths when the number of parameters is small?

\bibliography{ipo}
\end{document}